\documentclass[12pt]{article}
\usepackage{epsf,amsmath}
\begin{document}

\title{
EINSTEIN-SEPARABILITY, TIME RELATED HIDDEN PARAMETERS FOR
CORRELATED SPINS, AND THE THEOREM OF BELL }

\author{Karl Hess$^1$ and Walter Philipp$^2$}

\date{$^1$ Beckman Institute, Department of Electrical Engineering
and Department of Physics,University of Illinois, Urbana, Il 61801
\\ $^{2}$ Beckman Institute, Department of Statistics and Department of
Mathematics, University of Illinois,Urbana, Il 61801 \\ }
%\date{\today}
\maketitle

\begin{abstract}
We analyze the assumptions that are made in the proofs of
Bell-type inequalities for the results of Einstein-Podolsky-Rosen
type of experiments. We find that the introduction of time-like
random variables permits the construction of a broader
mathematical model which accounts for all correlations of
variables that are contained in the time dependent parameter set
of the backward light cone. It also permits to obtain the quantum
result for the spin pair correlation, a result that contradicts
Bell's inequality. Two key features of our mathematical model are
(i) the introduction of time operators that are indexed by the
measurement settings and appear in addition to Bell's source
parameters and (ii) the related introduction of a probability
measure for all parameters that does depend on the analyzer
settings. Using the theory of B-splines, we then show that this
probability measure can be constructed as a linear combination of
setting dependent subspace product measures and that the
construction guarantees Einstein-separability.

\end{abstract}
\medskip
\medskip

%\pacs{}
%
%\vskip2pc]
%

%%%%%%%%%%%%%%%%%%%%%%%%%%%%%%%%%%%%%%%%%%%%%%%%%%%%%%%%%%%%%%%%%%%%%%%%%%%%%
\section{INTRODUCTION}

We address the question whether the quantum result for the spin
pair correlation in Einstein-Podolsky-Rosen (EPR)-type experiments
can be obtained by a hidden parameter theory and show that it can
in spite of the serious objections given in the work of Bell
\cite{bell}.

The work of Bell \cite{bell} attempts to show that a mathematical
description of EPR-type experiments \cite{mermin},\cite{leggett}
by a statistical (hidden) parameter theory \cite{einstein} is not
possible. In EPR experiments, two particles having their spins in
a singlet state are emitted from a source and are sent to spin
analyzers at two spatially separated stations, $S_1$ and $S_2$.
The spin analyzers are described by Bell using unit vectors ${\bf
a},{\bf b}, etc.$ of three dimensional Euclidean space and
functions $A = \pm 1$ (operating at station $S_1$) and $B = \pm 1$
(operating at station $S_2$): furthermore A does not depend on the
settings $\bf b$ of station $S_2$, nor B on the settings $\bf a$
of station $S_1$ (Einstein separability or locality). Bell permits
particles emitted from the source to carry arbitrary hidden
parameters $\lambda$ of a set $\Omega$ that fully characterize the
spins and are "attached" to the particles with a probability
density $\rho$ (we denote the corresponding probability measure by
$\mu$) that does not depend on the settings at the stations. Bell
then assumes that the values of the functions A and B are
determined by the spin analyzer settings and parameters such that:
\begin{equation}
A = A({\bf a},{\lambda}) = {A_{\bf a}}(\lambda) = \pm 1
\label{eq1}
\end{equation}
and
\begin{equation}
B = B({\bf b},{\lambda}) = {B_{\bf b}}(\lambda) =\pm 1 \label{eq2}
\end{equation}
Thus ${A_{\bf a}}(\lambda)$ and ${B_{\bf b}}(\lambda)$ can be
considered as stochastic processes on $\Omega$, indexed by the
unit vectors $\bf a$ and $\bf b$ respectively. Quantum theory and
experiments show that, for a given time of measurement for which
the settings are equal in both stations, we have for singlet state
spins
\begin{equation}
{A_{\bf a}}(\lambda) = -{B_{\bf a}}(\lambda)  \label{eq3}
\end{equation}
with probability one. He further defines the spin pair expectation
value $P({\bf a}, {\bf b)}$ by
\begin{equation}
P({\bf a}, {\bf b}) = {\int}_{\Omega} {A_{\bf a}}(\lambda){B_{\bf
b}}(\lambda){\rho}({\lambda})d{\lambda} = -{\int}_{\Omega} {A_{\bf
a}}(\lambda){A_{\bf b}}(\lambda){\mu}(d{\lambda})  \label{eq4}
\end{equation}
From Eqs.(\ref{eq1})-(\ref{eq4}), Bell derives his celebrated
inequality \cite{bell}
\begin{equation}
1 +  P({\bf b}, {\bf c}) \geq \mid{P({\bf a}, {\bf b})  - P({\bf a}, {\bf c}) }\mid
\label{eq5}
\end{equation}
and observes that this inequality is in contradiction with the
result of Quantum Mechanics:
\begin{equation}
P({\bf a}, {\bf b}) = -{\bf a}\cdot{\bf b} \label{sept1}
\end{equation}

The proof of Bell's inequality is based on the obvious fact that
for $x, y, z = {\pm} 1$ we have
\begin{equation}
|xz -yz| = |x - y| = 1 - xy  \label{f8eq1}
\end{equation}
Substituting $x = {A_{\bf b}}({\lambda})$, $y = {A_{\bf
c}}({\lambda})$, $z = {A_{\bf a}}({\lambda})$ and integrating with
respect to the measure $\mu$ one obtains Eq.(\ref{eq5}) in view of
Eq.(\ref{eq4}).

In subsequent work by Clauser-Holt-Horne-Shimony (CHHS)
\cite{chhs} Eqs.(\ref{eq1})-(\ref{eq3}) were relaxed to permit
\begin{equation}
|{A_{\bf a}}| \leq 1 \text{   ,   } |{B_{\bf b}}| \leq 1
\label{f8eq2}
\end{equation}
They also introduced inequalities that do not deal with equal
settings and are of the form
\begin{equation}
|{{A_{\bf a}}({\lambda}){B_{\bf b}}({\lambda}) + {A_{\bf
d}}({\lambda}){B_{\bf b}}({\lambda}) + {A_{\bf
a}}({\lambda}){B_{\bf c}}({\lambda}) - {A_{\bf
d}}({\lambda}){B_{\bf c}}({\lambda})}| \leq 2 \label{f8eq3}
\end{equation}
This inequality is based on the observation that for $max(|x|,
|y|, |u|, |v|) \leq 1$
\begin{equation}
|u||x + y| + |v||x-y| \leq {|x + y| + |x - y|} \leq 2
\label{f8eq4}
\end{equation}
Thus, from the vantage point of mathematics, the Bell and CHHS
inequalities are straight-forward consequences of the respective
set of hypotheses and assumptions that are imposed.

We present an analysis of these hypotheses and assumptions with
the intent to form a new basis for a mathematical model of
EPR-experiments. The starting point of our analysis is an
examination of the role of time in the characterization of the set
of measured data that are represented by the functions ${A_{\bf
a}}$ and ${B_{\bf b}}$ in Eqs.(\ref{eq1}) and (\ref{eq2}). In a
complete EPR-experiment the settings $({\bf a}, {\bf b})$ are
randomly changed. However, in order to evaluate $P({\bf a}, {\bf
b})$ for the purpose of checking the Bell inequality, three
settings ${\bf a}, {\bf b}, {\bf c}$ will have to be selected and
considered as being fixed and the times at which the measurements
are taken for the three relevant pairs $({\bf a}, {\bf b})$,
$({\bf a}, {\bf c})$ and $({\bf b}, {\bf c})$ will now have to be
considered as being random. Thus, if a time dependence exists, it
is certainly reasonable to allow for an additional stochastic
parameter $\omega$ related to time which drives the random
processes taking place in the two stations. This parameter is not
correlated to $\lambda$ which can, for example, derive its
randomness from current and voltage fluctuations at the source.

The essence of our approach is the introduction of setting and
station specific time-like parameters as well as time related
setting dependent parameters ${\lambda}_{\bf a}^*$ on one side and
${\lambda}_{\bf b}^{**}$ on the other, that co-determine the
functions $A,B$ in addition to the correlated source parameters
$\lambda$(all to be defined more precisely in section 2). We show
in detail in the next sections that time-like parameters cannot be
fully covered by Bell's nor any of the other proofs mentioned
above. We also show that these parameters lead in a natural way to
setting dependent probability measures for the parameters without
spooky action at a distance. This was not considered a possibility
for three major reasons. First, as discussed below, a single
product measure $\mu = {\mu_{\bf a}}\times{\mu_{\bf b}}$ (where
${\mu_{\bf a}}$ only depends on ${\bf a}$ and ${\mu_{\bf b}}$ only
depends on ${\bf b}$ which would guarantee Einstein-separability
of the stations) cannot lead to the quantum result of
Eq.(\ref{sept1}). Second, stochastic parameters from the stations
(although investigated \cite{nus}) where not included as
integration variables in the probability measure because it seemed
impossible to reconcile the fact
\begin{equation}
{A_{\bf a}}(parameters) = -{B_{\bf a}}(parameters) \label{wk30}
\end{equation}
with that of setting dependent statistical parameters in the
spatially separated stations. The main reasoning was that the
settings are changed rapidly in between the measurements and the
parameters $\lambda$ can therefore carry no information on the
actual settings at the time period of measurement. Which
information could then possibly lead to Eq.(\ref{wk30}) without
invoking spooky action at a distance between the stations? We show
that time-like parameters (derived from a global clock time for
both stations) can provide this information. The third reason is
the widely held belief that if $A$ depends on $\bf a$, $\lambda$
and ${\lambda}_{\bf a}^*$, then by enlarging the parameter space
one can rewrite $A$ as a function of $\bf a$ and $\lambda$ with an
enlarged set of parameters $\lambda$; and similarly for $B$. We
will give a counterexample to this third point in appendix 1.

While all of the above will be extensively discussed, we would
like to emphasize already at this point that Bell has introduced a
number of assumptions on time dependencies (the use of
Eqs.(\ref{eq3}) without regard to the time-disjoint measurements)
and a significant asymmetry in describing the spin properties of
the particles  and the properties of the measurement equipment.
The spins are described by arbitrarily large sets $\Lambda$ of
parameters, while the measurement apparatus is described by a
vector of Euclidean space (the settings). This is true to Bohr's
\cite{bohr} postulate that the measurement must be classical. Yet,
the measurement apparatus must itself in some form contain
particles with spins that then, if one wants to be
self-consistent, also need to be described by large sets of
parameters that are related to the setting ${\bf a}, {\bf b},{\bf
c}...$.

%%%%%%%%%%%%%%%%%%%%%%%%%%%%%%%%%%%%%%%%%%%%%%%%%%%%%%%%%%%%%%%%%%
It is the purpose of this paper to show that a properly chosen sum
of what we call setting dependent subspace product measures
(SDSPM) does not violate Einstein-separability and does lead to
the quantum result of Eq.(\ref{sept1}) while still always
fulfilling Eq.(\ref{eq3}). By this we mean the following. The
probability space $\Omega$ is partitioned into a finite number $M$
of subspaces ${\Omega}_m$
\begin{equation}
\Omega = {{\cup}_{m=1}^M} {{\Omega}_m} \label{k13}
\end{equation}
A product measure ${({\mu_{\bf a}}\times{\mu_{\bf b}})}_m$ is
defined on each subspace ${\Omega}_m$. This measure can be
extended to the entire space $\Omega$ by setting
\begin{equation}
{{({\mu_{\bf a}}\times{\mu_{\bf b}})}_m}({{\Omega}_j}) = 0 \text{
if   } j \neq m \label{k14}
\end{equation}
which we denote by the acronym SDSPM. The final measure $\mu$ is
then defined on the entire space $\Omega$ by
\begin{equation}
\mu = {{\sum}_{m=1}^M} {({\mu_{\bf a}}\times{\mu_{\bf b}})}_m
\label{k2}
\end{equation}
As a preview we remark that the choice of subspace is determined
by time and setting dependent operators that symbolize the laws of
physics. In this way we introduce time-like setting dependent
parameters that are in no way suspect of spooky action at a
distance. At the same time the subspace measure becomes setting
dependent through these time-like parameters in a natural way.
This is, of course, the key element to our approach and is
therefore discussed in great detail in the bulk of the paper,
particularly in sections 2.1 and 3.1.
%%%%%%%%%%%%%%%%%%%%%%%%%%%%%%%%%%%%%%%%%%%%%%%%%%%%%%%%%%%%%%%%%%

We will show in the remainder of this paper that Bell's
assumptions are too restrictive and need to be relaxed. Moreover,
that this relaxation leads to a natural use of a sum of SDSPM's
which, in turn, gives the quantum mechanical result of
Eq.(\ref{sept1}). The proof of this is rather involved since
Einstein-separability needs to be guaranteed which imposes
stringent requirements on the possible sum of product measures; in
particular on the joint density and even conditional distributions
of setting dependent parameters.

\section{PARAMETER SETS BEYOND BELL's}

Before discussing possible generalized parameter sets, we would
like to make a case that time $t$ cannot be included as just
another parameter in Bell-type proofs.

\subsection{The special role of time}

We state up front what we believe to be the basis for obtaining
the quantum result with the use of hidden parameters: The
functions $A$ and $B$ and the densities $\rho$ of hidden
parameters may both relate to time without any involvement of
spooky action at a distance. Time correlations, even setting
dependent ones, may exist in both stations without any suspicion
or hint of spooky action. The introduction of these correlations
through time leads then to a probability measure that can depend
on the settings at both stations although the functions $A,B$
depend only on the settings of the respective stations and the
parameters, now considered as random variables, are independent
when averages over long time periods are taken. A well known fact
of probability theory is at the foundation of this: random
variables may be conditionally dependent (e.g. for certain time
periods) while they are independent when no conditions are
imposed.

Bell type proofs permit any number and form of parameters as long
separate integrations can be performed over the respective
densities i.e. if the joint conditional densities equal the
product of the individual conditional densities. The introduction
of time-like parameters presents then a critical problem since
other parameters in the argument of the functions $A$ and $B$ may
depend on time. This happens in an enormous number of natural
physical situations.

The following example is designed to define more clearly what we
understand by the term "time-like parameters". These parameters
may actually include space-like labels such as the settings.
However, with respect to their correlations they are time-like,
just as two clocks in two stations show time-like correlations
even if some space like settings (e.g the length of the pendulum)
are adjusted separately in the stations. To be definite, assume
that two stations have synchronized clocks with the pointer of
each clock symbolized by a vector of Euclidean space and denoted
by ${\bf s}_1$ in station $S_1$ and by ${\bf s}_2$ in station
$S_2$. Adding the setting vectors in the respective stations, one
obtains setting dependent time-like and correlated parameters
${{\bf s}_1} + {\bf a}$ in station $S_1$ and ${{\bf s}_2} + {\bf
b}$ in station $S_2$. One can find a natural implementation of
this example by using gyroscopes in the two stations located on
the rotating earth. If such parameters affect the functions $A$
and $B$, then integration over time cannot be factorized and time
cannot be introduced in Bell-type proofs without difficulty. We
note, in passing, that the rotation of the earth poses also the
following problem. The quantum result for the spin pair
correlation $P({\bf a}, {\bf b}) = -{\bf a}{\cdot}{\bf b}$ is
invariant to (time dependent) rotations while the mathematical
operations performed in the proof of Bells theorem are not. Thus
rotational symmetry is violated in Bell-type proofs through the
factorization process without assessment of its consequences.

To demonstrate the existence of hidden parameters in principle, we
may permit any parameter set that can be generated involving $t$
and local setting dependent operators $O_{{\bf a},t}^1$ in station
$S_1$ and $O_{{\bf b},t}^2$ in station $S_2$. These local
operators may act on any parameters (or information) in the
respective stations to create new parameters. For example, if a
particle that carries the parameter ${\lambda}_1$ arrives from the
source within a time period characterized by $\omega$ in station
$S_1$, then the time operator can transform this parameter into a
new "mixed" parameter ${{\Lambda}_{{\bf a},t}^1}({{\lambda}^1},
{\omega})$. Recall from section1 that the actual time of
measurement which determines $\omega$ must assumed to be random
since the settings are randomly switched \cite{bellbook}. We
distinguish this random time period $\omega$ from the time index
in the time operator because that dependence on time may or may
not be random. A more specific way of thinking about these
operators is by imagining two computers in the two stations which
have synchronized internal clocks. These computers can run any
program to create new parameters out of the locally available
input. Of equal importance, they can also be used to evaluate
these parameters i.e. assign them a value of $\pm 1$. Both
processes, creation and evaluation, may depend on the respective
setting and may be correlated in time.

We summarize now why time has a special standing and cannot be
included as just another parameter in Bell-type proofs.

\itemize

\item
Time may enter in more than one way to influence the value of the
functions $A$ and $B$. It influences this value through $\omega$
the randomly picked time (or time interval) of measurement at
given settings. It also influences the value through the
evaluation programs (the time operators). One does not need to
restrict oneself to one time operator per station, several such
operators could be used in any sequence depending on any
information available at the station i.e. any information
available in the backward light cone of a given station at the
time of measurement. This means that the functions $A$ and $B$ are
permitted to have an extended variable list and several variables
may be time related. The settings $\bf a$ and $\bf b$, the time
periods $\omega$ and the time-dependent evaluation programs (or
time operators) all contribute and determine the parameter random
variables that appear in the argument of the functions $A$ and $B$
and in the joint density $\rho$ of these parameter random
variables at a given time.
\item
The source parameters $\lambda$ may also depend on time. It is
possible that, depending on the information contained in the
backward light cone, the parameters $\lambda$ as well as the
frequency with which certain $\lambda$'s occur will be different
during different time periods. Actually even the number of
parameters that influence the measurement can be different during
different time periods; at least as far as principle is concerned.
\item
The above facts present grave difficulties to Bell's proof because
there may exist correlations between all these time dependent
parameter random variables and in fact these correlations must
exist to guarantee ${A_{\bf a}} = -{B_{\bf a}}$. Thus the joint
density $\rho$ of these parameter random variables may now show
time correlations and therefore $\rho$ may itself depend on time.
In particular, separate integration over the parameter values
needs to be justified and, in fact, may not even be possible. We
will see that this also presents great difficulty for the proof of
d'Espagnat \cite{espa} and other proofs that are, mathematically,
slight generalizations or special cases of Bell's.
\item
The possibility of the dependence of the joint density $\rho$ on
time and the relationship of the functions $A,B$ to time and
settings, as discussed above, make it convenient to shift the
assumption of time dependence to setting dependence. As explained
in sections 4 and 5, this yields the possibility to let both the
probability measure $\mu$ and the density $\rho$ associated with
it depend on the settings without introducing spooky action at a
distance. The mathematical model will be presented in section 4.
The principle for this is as follows. The evaluation of parameters
by the functions $A$ and $B$ (i.e. the value of $\pm 1$ that is
assigned for a certain parameter list) and the frequency with
which certain parameters appear may be correlated in time
including time correlations to the other station which influence
the possible choices of time related parameters that can be made
there for any given setting . This time correlation may also, in
each station, depend on the respective settings. Physically the
possibility of such a general probability measure arises because
the station equipment changes the incoming information (setting
dependent time operators) and evaluates then the changed
information within the same process.

It is also important to realize that the computer evaluation
programs may even be regarded as the "actual" setting at a given
station. There is a setting that the experimenter controls and
that influences the choice of computer program. However, for
different times the evaluation that the program performs may be
different. If both the times and settings are equal in the two
stations, then the evaluation is the same (to guarantee that
${A_{\bf a}} = -B_{\bf a}$ as demanded by Eq.(\ref{eq3})). This
view has a great significance: the number of actual different
settings in any given station may be vastly greater than the
number of choices that can be made by the experimenter without the
experimenter knowing. In other words, the hidden parameters may
not only represent the instruction set that is sent out for
evaluation but also represent the evaluation programs (i.e. the
"actual" settings) themselves and are correlated in time. The
number of different settings that the experimenter sees or
believes to be involved may therefore be much smaller than the
actual number involved in the evaluation. Note that this gives a
certain symmetry to measurement equipment and incoming parameters
that is absent in the approach of Bell.

The important question to be answered is now whether the proofs of
Bell-type theorems still can be carried out in a mathematically
rigorous way using these generalized time related parameter random
variables. We show in the next section for four major proofs
\cite{bellbook} of locality inequalities that time related
parameters of the kind discussed above do not permit a logical
execution of the proofs. Before doing so, we develop a more
precise mathematical notation and definition of possibly involved
time dependent parameter random variables.

\subsection{Definitions of time related parameters}

The starting point is a set of source parameters $\lambda =
(\lambda^1, \lambda^2)$ where the superscripts indicate
information carried to stations $S_1$ and $S_2$, respectively.
Random internal parameters $\lambda_{\bf {a}}^*$ operate at
station $S_1$ and $\lambda_{\bf {b}}^{**}$ at station $S_2$. In
other words, there is a layer of parameters below the mere
settings that will affect the values of the functions $A,B$. While
the observer might imagine that $A,B$ depend on settings only, the
values of the functions $A,B$ are determined by stochastic
processes, indexed by the unit vectors $\bf {a}$ and $\bf {b}$
respectively. For given vectors $\bf {a}$ and $\bf {b}$ we denote
the joint distribution of the resulting random variables
$\lambda_{\bf {a}}^*(\cdot)$ and $\lambda_{\bf {b}}^{**}(\cdot)$
by $\gamma = \gamma_{{\bf a}{\bf b}}$ which we allow to depend on
$\bf {a}$ and $\bf {b}$ in order to accommodate as broad a
situation as possible. A reasonable, though not a necessary,
assumption on $\gamma$ is the following continuity condition: for
fixed $\bf {a}$
\begin{equation}
  \lim_{{\bf b}\rightarrow{\bf a}} \gamma_{{\bf a}{\bf b}}
  \{ \omega: \lambda_{\bf {b}}^{**}(\omega) =
  \lambda_{\bf {a}}^*(\omega) \} \rightarrow
  1 = \gamma_{{\bf a}{\bf a}} \{ \omega:
  \lambda_{\bf {a}}^{**}(\omega) =
  \lambda_{\bf {a}}^*(\omega) \}.
\label{kleineq1}
\end{equation}
Intuitively speaking Eq.~(\ref{kleineq1}) says that if the vector
$\bf {b}$ at station $S_2$ is parallel or close to parallel to the
vector characterizing the analyzer setting in station $S_1$, then
for an ``overwhelming majority of cases $\omega$'' the
corresponding parameters $\lambda^*$ and $\lambda^{**}$ are equal.

Our probability space $\Omega$ consists of all pairs $(\lambda,
\omega)$, where $\lambda$ is a source parameter and $\omega$ is
the element of randomness related to time and driving the station
parameters $\lambda_{\bf {a}}^*(\cdot)$ and $\lambda_{\bf
{b}}^{**}(\cdot)$, respectively.  Furthermore, we assume that the
source parameters $\lambda$ will interact with the station
parameters $\lambda_{\bf {a}}^*(\omega)$ and $\lambda_{\bf
{b}}^{**}(\omega)$ with built in time $t$ dependence to form the
``mixed'' parameters $\Lambda_{{\bf {a}},t}^1(\lambda, \omega)$
and $\Lambda_{{\bf {b}},t}^2(\lambda, \omega)$ (one could
visualize this by some many body interactions). These are not free
parameters, but rather stochastic processes, indexed by the pairs
$({\bf {a}},t)$ and $({\bf {b}},t)$ at stations $S_1$ and $S_2$,
respectively, and defined on $\Omega$. The transition from
$(\lambda, \omega)$ to $\Lambda_{{\bf a},t}^1(\lambda, \omega)$
and $\Lambda_{{\bf b},t}^2(\lambda, \omega)$ is thought to be
defined by certain rules that can be represented by station
specific operators $O_{{\bf a},t}^1$ and $O_{{\bf b},t}^2$ that
depend on the globally known time $t$ that is the same at the
stations as well as at the source. Notice also that the time
operations and mixing of parameters occur during the collapse of
the wave-function (in quantum mechanical terms). The timing in
left and right stations and the values of time involved in the
measurement process are also quite flexible. It only needs to be
guaranteed that one deals with the same correlated pair.

Thus the connection between the time operators $O$ and the mixed
parameters $\Lambda$ is given by
\begin{equation}
  O_{{\bf a},t}^1({{\lambda}^1}, \omega) =
  O_{{\bf a},t}^1({{\lambda}^1}, \omega; \lambda_{{\bf a}}^*(\omega)) =
  \Lambda_{{\bf a},t}^1({{\lambda}^1}, \omega)
\label{kleineq2}
\end{equation}
and
\begin{equation}
  O_{{\bf b},t}^2({{\lambda}^2}, \omega) =
  O_{{\bf b},t}^2({{\lambda}^2}, \omega; \lambda_{{\bf b}}^{**}(\omega)) =
  \Lambda_{{\bf b},t}^2({{\lambda}^2}, \omega).
\label{kleineq3}
\end{equation}
Furthermore, the stochastic processes $A_{{\bf a},t}$ and $B_{{\bf
b},t}$ satisfy
\begin{equation}
  A_{{\bf a},t}(\lambda^1, \omega; \lambda_{{\bf a}}^*(\omega),
  \Lambda_{{\bf a},t}^1({{\lambda}^1}, \omega)) = \pm 1
\label{kleineq4}
\end{equation}
and
\begin{equation}
  B_{{\bf b},t}(\lambda^2, \omega; \lambda_{{\bf b}}^{**}(\omega),
  \Lambda_{{\bf b},t}^2({{\lambda}^2}, \omega)) = \pm 1,
\label{kleineq5}
\end{equation}
and if ${\bf b} = {\bf a}$ then in analogy to Eq.~(\ref{eq3}) we
have with probability $1$
\begin{equation}
  B_{{\bf a},t}(\lambda^2, \omega; \lambda_{{\bf a}}^{**}(\omega),
  \Lambda_{{\bf a},t}^2({{\lambda}^2}, \omega)) =
  -A_{{\bf a},t}(\lambda^1, \omega; \lambda_{{\bf a}}^*(\omega),
  \Lambda_{{\bf a},t}^1({{\lambda}^1}, \omega)).
\label{kleineq6}
\end{equation}
This means that the time operations for equal settings need to be
synchronized in order to lead to Eq.(\ref{kleineq6}). The
synchronization may be achieved by the selection of which settings
are chosen to be equal in the two stations and by the fact that
the stations are in the same inertial frame with identical
clock-time. (Time shifts and asymmetric station distances can
easily be accommodated in our model.)
%%%%%%%%%%%%%%%%%%%%%%%%%%%%%%%%%%%%%%%%%%%%%%%%%%%%%%%%%%%%%%%%

Assume therefore with us that station and setting dependent
parameter random variables influence EPR-type experiments. They
may be arbitrarily complicated mathematical objects. In the
simplest cases each parameter could, for example, be a matrix or
an n-dimensional vector etc.. We also may assume arbitrarily
complicated time operators that influence these parameters.

\section{TIME-LIKE PARAMETERS IN PROOFS OF BELL'S THEOREM}

We have selected the following proofs because they are
representative for all proofs of Bell's theorem that are known to
us and are described in Bell's book \cite{bellbook}. Before we
present them, we would like to clearly define the starting point
for the proofs. This starting point is the set of measured data
which is represented by the functions $A$ and $B$ as defined
(using our generalization) in Eqs.(\ref{kleineq4})
and(\ref{kleineq5}) with hidden parameters as also defined in this
equation.

\subsection{The Proof of Bell}

Bell \cite{bellbook} defines the following parameter sets that are
in the backward light cone (as defined by relativity). He lets $N$
denote the specification of all entities (called be-ables by Bell)
that are represented by parameters and belong to the overlap of
the backward light cones of both space-like separated stations
$S_1$ and $S_2$. In addition he considers sets of be-ables or
parameters $L_{\bf a}$ (our notation) that are in the remainder of
the backward light cone of $S_1$ and $M_{\bf b}$ for $S_2$
respectively. Bell then defines the conditional probability that
the function $A_{\bf a}$ assumes a certain value with $|A_{\bf a}|
\leq 1$ (this is a generalization of $A_{\bf a} = \pm 1$)
\begin{equation}
\{A_{\bf a}|{L_{\bf a}}, N\} \label{jan1}
\end{equation}
and similarly for $B_{\bf b}$
\begin{equation}
\{B_{\bf b}|{M_{\bf b}}, N\} \label{jan2}
\end{equation}
Then, to derive one of the celebrated "local inequalities" Bell
considers the expectation $E$ of the product ${A_{\bf a}}{ B_{\bf
b}}$:
\begin{equation}
E\{{A_{\bf a}}{B_{\bf b}}\} = {{\sum}_{AB}}{A_{\bf a}}{B_{\bf
b}}\{A_{\bf a}|{L_{\bf a}}, N\}\{B_{\bf b}|{M_{\bf b}}, N\}
\label{jan3}
\end{equation}

Here lies the crux of the problem with Bell-type derivations of
the locality inequalities. The parameter sets in the backward
light cones are not constant but evolve and are, certainly in
principle, different for all the different times at which each
single measurement is taken. In addition these parameters may be
time-like as far as their correlations are concerned. Using a more
precise notation one must therefore label the sets of be-ables
with indices that represent time $t$ (or time periods $\omega$)
e.g. by $N_t$, ${L_{{\bf a},t}}$ and ${M_{{\bf b},t}}$. Then it is
obvious that the summations above cannot be performed in a
straightforward fashion. However, to fully show that Bell's proof
does not go forward using the time dependent parameter space, one
must demonstrate that time cannot be just entered as another
parameter in Bell's proof (e.g. substituting $t$ for $\lambda$).
This demonstration is indeed possible because, as mentioned
frequently before, time may enter the functions $A,B$ and density
$\rho$ in form of two or more different variables. It is these
time-like variables, some indexed by the settings, that prevent
Bell-type proofs to go forward.

For example, one variable may be obtained by dividing the
time-axis in the frame of reference of the stations into equal
intervals, that are then randomly selected through the actual time
of measurement. We have denoted these time intervals by the
variable $\omega$. A second variable related to time may be
obtained by partitioning the time axis in station $S_1$ into a set
of intervals ${{\Delta}t}_{\bf a}$ that depend on the setting $\bf
a$ which, in turn, is randomly selected by the observer.

Define now $A_{\bf a}$ and $B_{\bf b}$ in two stations to be
functions of these different time related variables and also of an
additional variable $\lambda$ i.e.
\begin{equation}
A_{\bf a} = {A_{\bf a}}(\omega, {{\Delta}t}_{\bf a}, \lambda)
\label{bb8eq1}
\end{equation}
and
\begin{equation}
B_{\bf b} = {B_{\bf b}}(\omega, {{\Delta}t}_{\bf b}, \lambda)
\label{bb8eq2}
\end{equation}
We also define a probability density $\rho$
\begin{equation}
\rho = {\rho}(\omega, {{\Delta}t}_{\bf a}, {{\Delta}t}_{\bf b},
\lambda) \label{bb8eq3}
\end{equation}
The expectation of the spin pair correlation is then in Bell's
notation:
\begin{equation}
P({\bf a}, {\bf b}) = {\int}{A_{\bf a}}(\omega, {{\Delta}t}_{\bf
a}, \lambda){B_{\bf b}}(\omega, {{\Delta}t}_{\bf b},
\lambda){\rho}(\omega, {{\Delta}t}_{\bf a}, {{\Delta}t}_{\bf b},
\lambda) d{\omega} d{{\Delta}t}_{\bf a} d{{\Delta}t}_{\bf b}
d{\lambda} \label{bb8eq4}
\end{equation}
In the proofs of Bell-type theorems (also in the variation of
Clauser-Holt-Horne-Shimony \cite{chhs}) time-like parameters are
never explicitly considered. Station specific parameter random
variables, that may be denoted by ${\lambda}_{\bf a}^{*}$ in $S_1$
and by ${\lambda}_{\bf b}^{**}$ in $S_2$, are assumed to be
independent (considered as random variables). Bell-type proofs
contain then the following equation that is considered equivalent
to the validity of Einstein separability or locality for given
settings $\bf a$ and $\bf b$:
\begin{equation}
{\rho}({\lambda}_{\bf a}^{*}, {\lambda}_{\bf b}^{**}, {\lambda}) =
{{\rho}_1}({\lambda}){p_1}({\lambda}_{\bf
a}^{*}|{\lambda}){p_2}({\lambda}_{\bf b}^{**}|{\lambda})
\label{bb8eq5}
\end{equation}
where ${p_1}({\lambda}_{\bf a}^{*}|{\lambda})$ and
${p_2}({\lambda}_{\bf b}^{**}|{\lambda})$ denote conditional
probability densities given $\lambda$ and the settings $\bf a$ and
$\bf b$.

With the time-like variables as defined for Eq.(\ref{bb8eq4}),
Eq.(\ref{bb8eq5}) reads:
\begin{equation}
{\rho}(\omega, {{\Delta}t}_{\bf a}, {{\Delta}t}_{\bf b}, \lambda)
= {{\rho}_1}({\lambda}){{\rho}_2}({\omega}){p_1}({{\Delta}t}_{\bf
a}|{\lambda},{\omega}){p_2}({{\Delta}t}_{\bf b
}|{\lambda},{\omega}) \label{bb8eq6}
\end{equation}
However, for this equation to hold, the choices of time intervals
${{\Delta}t}_{\bf a}, {{\Delta}t}_{\bf b}$ would have to be
conditionally independent given the time interval $\omega$ and the
source parameter $\lambda$; but this cannot be the case, as they
are all correlated, i.e. connected with each other. Nor do
locality conditions have any consequence for the choice of
time-like intervals, even though they are indexed by the settings
in the respective stations, as long as the choices of the settings
are made separately and independent of the other station. Of
course, the dependence on time is "spatially non local" in the
sense that two clocks in separate stations may be perfectly
correlated at least when they are in the same inertial frame.

Some may be uncomfortable with the above choice of variables since
it is not easy to think of a general physical mechanism that, in
two stations, depends on various time intervals in selected ways.
We have therefore chosen in the main part of the paper time
operators $O_{\bf a}^1$ instead of the ${{\Delta}t}_{\bf a}^i$.
These time operators act on station specific and source specific
parameters (having in mind a simulation of many body interactions)
and result in the mixed parameters ${\Lambda}_{\bf a}(O_{\bf a}^1,
\omega, \lambda)$ in station $S_1$ and ${\Lambda}_{\bf b}(O_{\bf
b}^2, \omega, \lambda)$ in $S_2$. The fulfillment of physical
locality conditions in presence of time correlations is then not
as easy to show. However,  we do not invoke spooky action at a
distance as we will demonstrate in section 5.2. The core of this
demonstration is the mathematical fact that parameters in two
stations may be conditionally dependent (e.g. during certain time
periods) and simultaneously independent when no conditions are
imposed.

\subsection{The proof of Mermin}

Mermin's proof \cite{mermin} of the Bell inequalities was aimed at
a broad audience and considers only the essential basis and
consequences of Bell's theorem for a specific case that can be
experimentally realized. Three possible different settings where
assumed to be available to the experimenter in each of the two
stations ${S_1}, S_2$. The settings are chosen such that (slightly
paraphrased and transformed to agree with our notation):

(i) If one examines only those runs in which the settings are the
same in both stations, then the sign of the functions $A$ and $B$
is always opposite.

(ii) If one examines all runs without regard to what the settings
are, then one finds that the pattern of signs of $A$ and $B$ is
completely random. In particular, half the time the signs are the
same, and half the time different.

Since we have 2 different signs and three different settings on
each side, there are $2^3 = 8$ possible instructions that can be
given to determine the signs for the $3^2 = 9$ different and
random settings. It is then easy to see that the same signs must
occur in both stations $\frac {5} {9}$ of the time. This, however,
is in contradiction to requirement (ii). Mermin concludes
therefore that no instruction sets can exist.

However, from the discussion in the previous section we know that
the number of settings that the experimenter controls may be much
smaller than the "actual" hidden number of settings. In the
example of section 2.1, a different computer evaluation program
(that can be regarded as the actual setting) may be realized for
each different time of measurement and a given setting $\bf a$
chosen by the experimenter. If $N_s$ different computer evaluation
programs are realized for each of the different settings that the
experimenter can choose then we have $2^{3N_s}$ possible
instructions that can be given to determine the signs of the
${(3N_s)}^2$ different settings. Most importantly, however, only
the settings that the experimenter chooses are random. The
computer programs that form a "layer" below the chosen settings
may have time correlations among each other. The reasoning of
Mermin is based on the completely random choice of the settings.
This, however, is no longer guaranteed within the framework of our
time related and time-like parameters that are represented by the
computer programs. These arguments present also difficulties for
the proof of d'Espagnat \cite{espa}. In defense of Bell one could
say that such correlations must also run into problems with spooky
action. How can (ii) be true if time correlations exist
"underneath" the random switching of the experimenter. We will
show, however, that this suspicion has no mathematical basis. As
mentioned in section 2.1 random variables may be conditionally
dependent (e.g. for certain time periods) while they are
independent when no conditions are imposed.  In other words one
can have correlations through certain time periods and for certain
given source parameters $\lambda$ without the necessity of
correlations when no constraint is applied e.g. over long periods
of time. Note also that we have only used local operators and
there is no suspicion of spooky action as long as we have no
global (unconditional) correlations of space like parameters.
There may be correlations in time that extend over space.

Mermin's proof of a Bell type inequality can therefore not proceed
when time-like parameters and setting dependent (within a station)
time operators are involved.

\subsection{ Variations of the proof of d'Espagnat}

Some variations of d'Espagnats proof \cite{espa} appear at first
glance different to that of Bell although they form a special case
by substituting for the general measure (considered in Bell's
proof) a properly normalized sum of point masses. We give below a
prototype version of such proofs.

Assume we have chosen the setting vectors ${\bf a}, {\bf d}$ in
station $S_1$ and setting vectors ${\bf b}, {\bf c}$ in station
$S_2$. Further assume that we can rearrange all measurements in
such a way that they can be concatenated in groups of four that
then fulfill the following inequality
\begin{equation}
|{{A_{\bf a}}(...){B_{\bf b}}(...) + {A_{\bf d}}(...){B_{\bf
b}}(...) + {A_{\bf a}}(...){B_{\bf c}}(...) - {A_{\bf
d}}(...){B_{\bf c}}(...)}| \leq 2 \label{bb3eq1}
\end{equation}
with $(...)$ denoting a certain subset of the parameters. Assume
further that the union of all subsets $(...)$ gives all parameters
that can possibly describe the given set of experiments. We call
the statement of this paragraph the rearrangement assumption (RA).

The inequality of Eq.(\ref{bb3eq1}) follows from Eq.(\ref{f8eq4}).
Averaging over all parameters $(...)$, one obtains the spin pair
expectation values for the various settings. The averaged
Eq.(\ref{bb3eq1}) represents then an important "locality
inequality" of which Bell's is a special case.

Note, that this type of proof starts from the RA and the
inequality of Eq.(\ref{bb3eq1}). Implying that all the
measurements can be covered that way if parameters are inserted,
one arrives at the inequality for the average spin-pair
correlation. The explanation offered to show that indeed all
measurements can be covered that way is based on (i) the avoidance
of spooky action (AoSA), (ii) some form of inductive logic (IL)
based on the fact that repeated experiments with reasonably large
numbers of measurements must give about the same result.

From (i) it is deduced that the list of parameters that appears in
the arguments of the functions $A$ and $B$ contains all possible
combinations of all possible parameters, independent of the
particular setting. The reasoning is approximately like this: if
the parameters that appear in the arguments of $A$ and $B$ would
depend on the setting, then by switching from one setting to the
other in station $S_1$ something must happen to the parameter set
in station $S_2$ which would be spooky action. This reasoning,
however, is (as in the proof of Mermin) not mathematically sound.
We repeat that it is well known in probability theory that random
variables may be conditionally dependent but independent when no
conditions (or different ones) are imposed. This opens the
possibility that there exist different conditional dependencies
for the various settings while the parameters are independent when
viewed without condition. For example, for time periods during
which certain time operators are at work and/or certain parameters
$\lambda$ are emitted from the source, the parameters in station
$S_1$ may be correlated to those in station $S_2$ in other words
are conditionally dependent. The different conditional
dependencies mean that the listings of the parameters do not
contain all possible combinations of the parameters for all
settings and for all time intervals.

One might think that inductive logic (point (ii)) may save this
type of proof. However, long term averages can still be the same
in spite of the presence of time correlations. They are guaranteed
to be the same in our model.

We summarize these ideas by presenting our argument in a slightly
different way. Clearly, time cannot be partitioned into a finite
number of elements that randomly repeat themselves. We also note,
that for each particular setting and measurement the time interval
$\omega$ may be different. Let's enumerate then the time intervals
in the measurements with setting ${\bf a}, {\bf b}$ by
${{\omega}_j}$ and with setting ${\bf d}, {\bf b}$ by
${{\omega}_{j*}}$ with $j \neq {j*}$ since the measurements must
necessarily be at different times. The first two terms of
Eq.(\ref{bb3eq1}) are then
\begin{equation}
{A_{\bf a}}(.{{\omega}_j}.){B_{\bf b}}(.{{\omega}_j}.) + {A_{\bf
d}}(.{{\omega}_{j*}}.){B_{\bf b}}(.{{\omega}_{j*}}.)
\label{bb3eq2}
\end{equation}

The rearrangement assumption (RA) means that all parameters must
appear in all arguments of the functions $A$ and $B$ independent
of setting and that therefore a reordering is possible to obtain
Eq.(\ref{bb3eq1}). But how can that be proven? It is not even
necessary that there exists a periodic repetition of parameters.
The parameters that are available at each time of measurement in
each station comprise all the information up to present that is
contained in the backward light cone \cite{bellbook}. Since the
backward light cone is $\bf {different}$ for each different
measurement a different parameter or different combination of
parameters can be selected (at least in principle) each time. The
above proof needs to assess then the properties and functional
dependencies of the possible parameters ${{\Lambda}_{{\bf
a},t}^1}({{\lambda}^1}, {\omega})$ and operators $O_{{\bf a},t}^1$
etc. which could permit a reordering into the sets of four shown
in Eq.(\ref{bb3eq1}) (for the case when the continuum of time is
involved and for virtually arbitrary time operators). There is no
proof in the literature known to us which does even address these
questions. The only arguments that are given start with
Eq.(\ref{bb3eq1}) and use induction in the backward direction of
the proof. Using small finite sets of parameters and
Eq.(\ref{bb3eq1}) one can proceed to larger and larger sets.
However, the proof needs to start from the diversity of
Eq.(\ref{bb3eq2}) and proceed to derive Eq.(\ref{bb3eq1})by
reordering. Of course, we cannot directly show that this is not
possible. However, if one can find a local set of parameters that
gives the quantum result (as we believe we have below), then all
the variations of  d'Espagnat's proof are refuted.

\subsection{The Proof of Clauser-Holt-Horne-Shimony}

The proof of Clauser-Holt-Horne-Shimony (CHHS) \cite{chhs}
introduces a variation of Bell's inequality and permits a
violation of ${A_{\bf a}} = -{B_{\bf a}}$ i.e. of Eq.(\ref{eq3})
to any degree. In addition, the values of the functions $A$ and
$B$ can be such that $|A|, |B| \leq 1$ . This violation is caused
by station specific setting dependent parameters that have no
correlation from one station to the other. Note that therefore
these station specific parameters (see \cite{chhs} and also
\cite{nus}) are very different from the ones introduced by us. Our
station specific parameters are correlated by clock time and do
neither lead to any violation of Eq.(\ref{eq3}) nor to absolute
values of the functions $A$ and $B$ that are smaller than $1$.

The advantages of CHHS lie in the fact that their inequality does
not contain exactly equal settings that are experimentally
difficult to achieve. They use instead the inequality shown
already above
\begin{equation}
|{{A_{\bf a}}(...){B_{\bf b}}(...) + {A_{\bf d}}(...){B_{\bf
b}}(...) + {A_{\bf a}}(...){B_{\bf c}}(...) - {A_{\bf
d}}(...){B_{\bf c}}(...)}| \leq 2 \nonumber
\end{equation}
They also have a natural explanation for the experimentally
observed deviations from ${A_{\bf a}} = -{B_{\bf a}}$: random
influences of parameters or any type of fluctuations at the
stations similar to noise. However, these deviations come at the
price of also violating the quantum result. Their random
fluctuations at the stations, when fully effective, will
completely destroy the quantum result. Certainly, the Bell
inequalities (or CHHS inequalities) stay valid. This is, however,
without logical consequence since under these circumstance it is
clear that the quantum result will not be experimentally
confirmed. If the noise is weak, then part of the quantum result
is recovered. However, only to the extent that ${A_{\bf a}} =
-{B_{\bf a}}$. We need to consider therefore, as far as principle
is concerned, only the subset of measurements and parameters for
which ${A_{\bf a}} = -{B_{\bf a}}$ may be implied. For this subset
the CHHS inequality brings nothing new. Replace above the settings
${\bf c}{\rightarrow} -{\bf c}$ and ${\bf d}{\rightarrow} {\bf c}$
and the original Bell inequality of Eq.(\ref{eq5}) is recovered.
Therefore, CHHS need not be treated separately from the proof of
Bell as far as the principle and the fundamental deviations from
the quantum result are concerned. We emphasize, however, that our
station specific parameters are very different from those of CHHS.
Ours are time correlated and do not violate ${A_{\bf a}} =
-{B_{\bf a}}$.

\section{A THEOREM FOR HIDDEN EPR-PARAMETER SPACES}

Our goal is to show that under our generalized conditions, it is
possible to obtain the quantum result, the scalar product $-{\bf
a}\cdot{\bf b}$ for the spin pair expectation value $P({\bf
a},{\bf b})$. This task will be completed in several installments
in the next section. Here we formulate a theorem which provides
the stepping stone for this procedure. Note that our measure
deviates from a probability measure by at most $\epsilon$, which
can be chosen arbitrarily small. We believe that this presents no
physical limitation of the theory but include it for reasons of
mathematical precision.

\noindent $\bf {THEOREM }$: Let $0<{\epsilon}<\frac {1} {2}$ and
let ${\bf a} = (a_1, a_2, a_3)$ and ${\bf b} =(b_1, b_2, b_3)$ be
unit vectors. Then there exists a finite measure space $(\Omega,
{{\bf F}}, \mu = {\mu}_{{\bf a},{\bf b}})$ and two measurable
functions $A$ and $B$ defined on it with the following properties:
\begin{equation}
{\Omega}\subset{R^2} \text{and {\bf F} depend on  } {\epsilon}
\text{ only} \label{wk1}
\end{equation}
$\Omega$ is a compact set. Its elements are denoted by $(u,v)$.

The measure $\mu$ only depends on ${\bf a}, {\bf b}$ and
$\epsilon$, satisfies
\begin{equation}
1 \leq {{\mu}({\Omega})}< 1 + {\epsilon} \label{wk2}
\end{equation}
and has a density ${\rho}_{{\bf a}{\bf b}}$ with respect to
Lebesgue measure.

The functions $A$ and $B$ assume the values
\begin{equation}
A,B = \pm1 \label{wk4}
\end{equation}
and $A$ depends only on $\bf a$ and $u$, B only on $\bf b$ and
$v$.

Further
\begin{equation}
E\{{A_{\bf a}}{B_{\bf b}}\} = {\int}_{\Omega} {A_{\bf
a}}(u){B_{\bf b}}(v){{\rho}_{{\bf a}{\bf b}}}(u,v) dudv = -{\bf
a}\cdot{\bf b} \label{wk5}
\end{equation}

and for each vector $\bf a$ the following equation holds for all
$(u, v)\in{\Omega}$ except on a set of $\mu$-measure $< \epsilon$:
\begin{equation}
{B_{\bf a}}(u,v) = -{A_{\bf a}}(u,v) \label{wk6}
\end{equation}
(Note that mathematical precision requires the listing of all the
parameters in the functions so that the integral of Eq.(\ref{wk5})
is well defined. However, since it is important for the physics
that the functions actually depend only on a subset of parameters,
as mentioned after Eq.(\ref{wk4}) and as will become clear when
the meaning of $u, v$ is fully described, we list in the arguments
of the functions only that subset.)

The proof of the theorem requires the following fact which follows
from a basic theorem on B-splines \cite{schumaker}. We state the
fact here in form of a lemma.

\noindent$\bf {Lemma}$: Let $n \geq 4$ be an integer. Then there
exist real-valued functions ${N_i}(x),{{\psi}_i}(y)$ with $1 \leq
i \leq n$ depending on real variables $x$ and $y$, respectively,
such that
\begin{equation}
{0 \leq {{N_i}(x)} \leq 1}\text{ , } {0 \leq {{\psi}_i(y)} \leq 2}
\text{  for } 0 \leq {x,y} \leq 1 \label{wk7}
\end{equation}
and
\begin{equation}
0 \leq {{\sum}_{i=1}^n}{{\psi}_i}(y){N_i}(x) - {{(y-x)}^2} \leq
{{\frac {1} {4}}{n^{-2}}}\text{ for } 0 \leq {x,y} \leq 1
\label{wk8}
\end{equation}
The proof for this lemma is given in appendix 2. We now proceed to
prove the main theorem.

\noindent$\bf {Proof}$ of $\bf {Theorem}$: Choose an even integer
$n>1/\epsilon$ and for $\Omega$ the square $\Omega =
{{[-3,3n)}^2}$ with side of length $3 + 3n$. We endow $\Omega$
with Lebesgue measurability, symbolized by the $\sigma$-field
${\bf F}$ and define:

\begin{eqnarray}
{A_{\bf a}}(u) &=& \left \{
\begin{array} {cll}
{\text{sign}(a_k)} & \mbox{if $-k \leq u  < -k+1$} & k=1,2,3 \\

-1 & \mbox{if $2j  \leq u  < 2j+\frac{1}{2}$} &
j=0,1,...,\frac{3n}{2}-1 \\

+1 & \mbox{if $2j+\frac{1}{2} \leq u < 2j+2$} &
j=0,1,...,\frac{3n}{2} \\
\end{array}
\right.
\end{eqnarray}

Thus $A$ depends on ${\bf a}$ and $u$ only. Here and throughout we
set $\text{sign}(0)=1$. Similarly, we define

\begin{eqnarray}
{B_{\bf b}}(v) &=& \left \{
\begin{array} {cll}
{-\text{sign}(b_k)} & \mbox{if $-k \leq v  < -k+1$} & k=1,2,3 \\

+1 & \mbox{if $2j+\frac{1}{2}  \leq v  < 2j+\frac{3}{2}$} &
j=0,1,...,\frac{3n}{2}-1 \\

-1 & \mbox{if $2j-\frac{1}{2} \leq v < 2j+\frac{1}{2}$} &
j=0,1,...,\frac{3n}{2}-1 \\ -1 & \mbox{if $0 \leq v <
\frac{1}{2}$}\text{     or} & 3n-\frac{1}{2} \leq v <3n\\
\end{array}
\right.
\end{eqnarray}

Thus, $B$ depends on ${\bf b}$ and $v$ only. Notice that on
${{\cup}_{k=1}^3}{[-k,-k+1)}^2$, Eq.(\ref{wk6}) is satisfied for
{\em{all}} values of $(u, v)$.

Next, we define
\begin{equation}
\sigma_{\bf a}(u) = \label{ph1}
\end{equation}
\begin{eqnarray}
\nonumber
\begin{array} {ll}
|a_k| \cdot 1\{-k \leq u < -k+1\}  & k \in I_1
\\

N_k(|a_1|) \cdot 1\{{k-1} \leq u < k\}  & k \in I_2
\\

N_{k-n}(|a_2|) \cdot 1\{{k-1} \leq u < k\}  & k \in I_3
\\

N_{k-2n}(|a_3|) \cdot 1\{{k-1} \leq u < k\}  & k \in I_4
\\
\end{array}
\end{eqnarray}
\begin{equation}
\tau_{\bf b}(v) = \label{ph2}
\end{equation}
%------------------------------------------------------------
\begin{eqnarray}
\nonumber
\begin{array} {ll}
|b_k| \cdot 1\{-k \leq v < -k+1\}  & k \in I_1
\\

\frac{1}{2}\psi_k(|b_1|) \cdot 1\{{k-1} \leq v < k\} & k \in I_2
\\

\frac{1}{2}\psi_{k-n}(|b_2|) \cdot 1\{{k-1} \leq v < k\} & k \in
I_3
\\

\frac{1}{2}\psi_{k-2n}(|b_3|) \cdot 1\{{k-1} \leq v < k\} & k \in
I_4
\\

\end{array}
\end{eqnarray}
The symbols $I_1..I_4$ stand for: $I_1 = -3, -2, -1$; $I_2 =
1,...,n$; $I_3 = n+1,...2n$; $I_4 = 2n+1,...,3n$ and $1\{\cdot\}$
denotes the indicator function. Furthermore, let $$ \delta_{jk}
=
\left \{
\begin{array} {ll}
1 & \mbox{if $j=k$} \\ 0 & \mbox{if $j \neq k$} \\
\end{array}
\right. $$ be the Kronecker symbol. We set
%------------------------------------------------------------
\begin{equation}
\nu(u, v) =  \delta_{ij} \cdot 1\{i-1 \leq u < i\}{\cdot} 1\{j-1
\leq v < j\}\text{  } i,j=-2,-1,...,3n
\end{equation}
%------------------------------------------------------------

We finally define the density ${\rho}_{{\bf a}{\bf b}}$ by
\begin{equation}
{\rho}_{{\bf a}{\bf b}}(u, v) = {{\sigma}_{\bf a}}(u){{\tau}_{\bf
b}}(v){\nu}(u, v) \label{kw1}
\end{equation}
and the measure ${\mu}_{{\bf a}{\bf b}}$ by having density
${\rho}_{{\bf a}{\bf b}}$ with respect to Lebesgue measure. This
definition, of course, entails that ${\mu}_{{\bf a}{\bf b}}$ is a
sum of SDSPM's. The integrals that we have to perform will then
correspond to summations over integrals of such product measures.
Note that a single product measure of independent events would, of
course, assuage all concerns related to spooky action but cannot
yield the quantum result because the two spin measurements are not
independent (see also appendix 3). There are considerable
correlations possible because of the correlations of the source
parameters and because of the knowledge of clock time (the time
operator) in both stations. We therefore have introduced
correlations by partitioning the measure into a sum of SDSPM's. In
the above equations, the correlation is expressed by the Kronecker
symbols. Note, however, that the diagonal arrangement above is not
necessary and leads only to one particular sum of SDSPM's. A large
number of off-diagonal arrangements can also be included as we
will see below and the actual sum of product measures that we use
is a superposition of all these possibilities. This will enable us
to obtain a uniform joint density and thus to avoid any sign of
spooky action.

From the above definitions we obtain the following integrals for
the spin pair correlation functions:
\begin{equation}
{{\int}_{{[-3,0)}^2}}{A_{\bf a}}(u) {B_{\bf b}}(v){\rho}_{{\bf
a}{\bf b}}(u, v)d u d v =
-{{\sum}_{k=1}^3}\mid{a_k}\mid\mid{b_k}\mid{sign ({a_k}) sign
({b_k})} = -{\bf a}\cdot{\bf b} \label{wk10}
\end{equation}
Furthermore, the integral over the complement of the square ${[-3,
0)}^2$ vanishes i.e.
\begin{equation}
{{\int}_{{\Omega}{\setminus}{[-3,0)}^2}}{A_{\bf a}}(u) {B_{\bf
b}}(v){\rho}_{{\bf a}{\bf b}}(u, v)d u d v = 0 \label{wk11}
\end{equation}
which proves Eq.(\ref{wk5}).

It remains to be shown that ${\rho}_{{\bf a}{\bf b}}$ defines a
measure $\mu$ that is close, within $\epsilon$, to a probability
measure i.e. fulfills Eq.(\ref{wk2}). For this, we consider the
mass distribution between the square ${[-3,0)}^2$ and its
complement. The amount of mass $M_1$ distributed over ${[-3,0)}^2$
is
\begin{equation}
M_1 = {{\sum}_{k=1}^3}\mid{a_k}\mid\mid{b_k}\mid \label{wk12}
\end{equation}
The mass $M_2$ of ${{\Omega}{\setminus}{[-3,0)}^2}$ equals
\begin{equation}
M_2 = {\frac {1}
{2}}{{\sum}_{k=1}^3}{{\sum}_{i=1}^n}{N_i}(\mid{a_k}\mid){{\psi}_i}(\mid{b_k}\mid)
\label{wk13}
\end{equation}
Thus the total mass distributed equals in view of Eq.(\ref{wk8})
\begin{equation}
M_1 + M_2 = {{\sum}_{k=1}^3}\mid{a_k}\mid\mid{b_k}\mid + {\frac{1}
{2}}{{\sum}_{k=1}^3}{{\sum}_{i=1}^n}{N_i}(\mid{a_k}\mid){{\psi}_i}(\mid{b_k}\mid)
\nonumber
\end{equation}
\begin{equation}
M_1 + M_2 = {{\sum}_{k=1}^3}\mid{a_k}\mid\mid{b_k}\mid + {\frac{1}
{2}} {{\sum}_{k=1}^3}{{(\mid{a_k}\mid - \mid{b_k}\mid)}^2} +
{\theta}\cdot{n^{-2}} \nonumber
\end{equation}
\begin{equation}
M_1 + M_2 =  1 + {\theta}\cdot{n^{-2}} < 1 + \epsilon \label{kw60}
\end{equation}
where $0 \leq \theta <1/24$. For the case ${\bf b} = {\bf a}$ we
have
\begin{equation}
0 \leq M_2 < {\theta}\cdot{n^{-2}} < \epsilon
\end{equation}
As was observed right after the definitions of $A$ and $B$,
Eq.(\ref{wk6}) holds for all
$(u,v)\in{{\cup}_{k=1}^3}{[-k,-k+1)}^2$ and thus for all $(u,
v)\in\Omega$ except, perhaps, on a set of $\mu$-measure
$<\epsilon$. This completes the proof of the theorem.

The proof clearly shows that for ${\bf b} = {\bf a}$ we can choose
$\Omega$ to be a probability space, i.e. $\mu(\Omega) = 1$.

\section{QUANTUM RESULT WITHOUT SPOOKY ACTION}

\subsection{Connection to EPR-experiments}

Suppose now that ${{\Lambda}_{{\bf a},t}^1}, {{\Lambda}_{{\bf
b},t}^2}$ are mixed parameters as defined above. Let $f$ and $g$
be real-valued bounded functions on the space of the
${{\Lambda}_{{\bf a},t}^1}$'s and ${{\Lambda}_{{\bf b},t}^2}$'s.
We do not assume that these two $\lambda$-spaces are identical,
nor is it necessary to specify them at this point. However, we
need to assume that, for fixed $\bf a$,$\bf b$ and time operators,
the mappings $f({{\Lambda}_{{\bf a},t}^1})$ and
$g({{\Lambda}_{{\bf b},t}^2})$ from $\Omega \rightarrow R$ are
measurable so that they can be considered as random variables.
Since $f$ and $g$ are assumed to be bounded, we may assume without
loss of generality that the ranges of $f({{\Lambda}_{{\bf
a},t}^1})$ and $g({{\Lambda}_{{\bf b},t}^2})$ equal the interval,
[-3,3n]. A mathematical model for EPR-experiments can now be
obtained by an application of the theorem. For fixed time
operators and source parameters $\lambda$ we define the joint
density of $f({{\Lambda}_{{\bf a},t}^1})$ and $ g({{\Lambda}_{{\bf
b},t}^2})$ to equal ${\rho}_{{\bf a}{\bf b}}(u, v)$, as defined in
Eq.(\ref{kw1}). Then by Eq.(\ref{wk5}) and by the standard
transformation formula for integrals we have for fixed $\lambda$
and time operators $O_{{\bf a},t}^1$, $O_{{\bf b},t}^2$:
\begin{equation}
E\{{A_{{\bf a},t}}({{\lambda}^1}, \cdot, f({{\Lambda}_{{\bf
a},t}^1}({{\lambda}^1}, \cdot))){B_{{\bf b},t}}({{\lambda}^2},
\cdot, g({{\Lambda}_{{\bf b},t}^2}({{\lambda}^2}, \cdot)))\} =
-{\bf a}\cdot{\bf b} \label{w6}
\end{equation}
Here the expectation $E$ operates on the space of $\omega$, a
subspace of $\Omega$; the dummy variable $\omega$ of the
integration is symbolized by $(\cdot)$.

This direct application does not address the key question whether
the introduced probability measure is free of the suspicion of
spooky action at a distance. To show this, we need to ensure the
following. If setting $\bf b$ at station $S_2$ is changed into
setting $\bf c$, the probability distribution governing the
parameters ${{\Lambda}_{{\bf a},t}^1}$ at station $S_1$ must
remain unchanged. The fact that the ratios of the relative
frequencies ${{\Lambda}_{{\bf a},t}^1}/{{\Lambda}_{{\bf b},t}^2}$
and ${{\Lambda}_{{\bf a},t}^1}/{{\Lambda}_{{\bf c},t}^2}$ may be
different is not of concern. The time operator defined above can
easily account for this. However, the average frequencies of the
parameters $f({{\Lambda}_{{\bf a},t}^1})$ in each of the intervals
between $-3$ and $3n$ must not change when setting ${\bf b}$ is
changed in the other station. This can be accomplished with ease
by superposition of variations of the above described element in
the following two step operation.

\subsection{Avoidance of spooky action}

Choose any of the ${(n + 1)}^2$ squares $Q_{jk}$ with vertices at
the points $(3j,3k), (3(j+1),3k)$, $(3(j+1), 3(k+1))$ and
$(3j,3(k+1))$ for $j,k = -1,0,1,2,...,n-1$. Now repeat the entire
construction with $Q_{jk}$ replacing $Q_{-1-1}$. Define $A$ and
$B$ to be equal to $sign({a_i})$ or $-sign({b_i})$, respectively,
on each of the three vertical and horizontal strips of $Q_{jk}$
with $i = 1,2,3$. On the vertical and horizontal strips not
containing parts of $Q_{jk}$ define $A$ and $B$ equal $\pm1$ in an
obvious modification of the above construction. More precisely, we
perform the following operations. As far as the definition of $A$
is concerned, we interchange the vertical strips
$[-3,0){\times}[-3,3n)$ and $[3j,3(j+1)){\times}[-3,3n)$.
Similarly, for the definition of $B$ we interchange the horizontal
strips $[-3,3n){\times}[-3,0)$ and $[-3,3n){\times}[3k,3(k+1))$.
Next assign mass $M_1$ to $Q_{jk}$  and mass $M_2$ to the
complement ${\Omega}\setminus{Q_{jk}}$ of $Q_{jk}$ . $M_2$ will be
distributed on $3n$ unit squares as follows: $Q_{jk}$ and the
vertical and horizontal strips associated with them take a total
of $(3n+3).3.2 -9 = 18n+9$ unit squares. From the remaining
$9{n^2}$ unit squares we choose $3n$ and distribute the mass
${\frac {1} {2}}{N_i}(\mid{a_k}\mid){{\psi}_i}(\mid{b_k}\mid)$ on
them (with $1 \leq i \leq n$ and $1 \leq k \leq 3$). For given
$Q_{jk}$ this yields
\begin{equation}
N = {(n + 1)^2} {\binom {9n^2} {3n}} \label{eqnov1}
\end{equation}
possible measures ${\mu}_m$ with $1 \leq m \leq N$. For each of
these measures Eqs.(\ref{wk1})-(\ref{wk6}) hold. Label the
corresponding functions $A$ and $B$ as $A_{(m)}$ and $B_{(m)}$ and
consider the index $(m)$ a function of the source parameter
$\lambda = ({{\lambda}^1}, {{\lambda}^2})$ and the time operators
$O_{{\bf a},t}^1$, $O_{{\bf b},t}^2$. With respect to the
dependence on time we make the usual assumptions, such as a
possible invariance with respect to certain translations. Notice
that variations (with settings) of the frequencies of setting
dependent parameters in certain given time intervals are permitted
by the properties of the time operator and do not indicate spooky
action. Then the functions $A_{(m)}$ and $B_{(m)}$ can be
considered as functions of $\bf a$, $\lambda$, ${{\Lambda}_{{\bf
a},t}^1}$, and $\bf b$, $\lambda$, ${{\Lambda}_{{\bf b},t}^2}$,
respectively. Finally define a new measure $\mu$ on $\Omega$ by
setting
\begin{equation}
\mu = {\frac {1} {N}}{{\sum}_{m=1}^{N}} {\mu}_m \label{kw65}
\end{equation}
At this point we consider $\Omega$ as the union of $N$ layers of
the above type stacked up in three dimensions, reinterpreting
$A$,$B$ and $\mu$ accordingly.

The second step in the modification of the construction is a minor
variation of the first one and, depending on ones taste, may not
be needed. Instead of lining up mass $M_1$ on the diagonal of the
squares $Q_{jk}$, we assign this mass to three unit squares within
$Q_{jk}$ such that each vertical and horizontal row contains
exactly one unit square with mass $\mid{a_i}\mid\mid{b_i}\mid,i =
1,2,3$. Moreover, these are permuted so that they yield  in $3! =
6$ ways yielding a total of $36$ possibilities. Taking these into
account we define now $\mu$ as the average of $36{(n+1)^2}{\binom
{9n^2} {3n}}$ measures ${\mu}_m$ as in Eq.(\ref{kw65}) but now
with $N$ denoting the total number of measures involved. Just as
before we note that $A$, $B$ and $\mu$ satisfy Eqs.(\ref{wk1})
through (\ref{wk6}) and that $\Omega$ is now defined as the union
of $N$ layers in three dimensions.

Let us inspect the density ${\rho}_{{\bf a}{\bf b}}$ which is now
defined on the domain
\begin{equation}
{\rho}_{{\bf a}{\bf b}}(u, v, m)\text{  with  } -3 \leq u,v < 3n;m
= 1,2,...,N. \label{kw67}
\end{equation}
For fixed $u$ and $v$, the joint density governing the pair of
parameters $f({{\Lambda}_{{\bf a},t}^1})$ and $g({{\Lambda}_{{\bf
b},t}^2})$ is given by
\[ = {\frac {1} {N}}{{\sum}_{m=1}^N} {\rho}_{{\bf a}{\bf b}}(u, v, m)
\]
\[ = {\frac {1} {{(3n+3)}^2}}({{\sum}_{k=1}^3}\mid{a_k}\mid\mid{b_k}\mid + {\frac{1}
{2}}{{\sum}_{k=1}^3}{{\sum}_{i=1}^n}{N_i}(\mid{a_k}\mid){{\psi}_i}(\mid{b_k}\mid))\]
\[ = {\frac {M_1 + M_2} {{(3n+3)^2}}} = {\frac {1 + \theta\cdot{n^{-2}}}
{{(3n+3)^2}}}\] with $\theta \leq 1/24$ in view of
Eq.(\ref{kw60}). This shows that the joint density of
$f({{\Lambda}_{{\bf a},t}^1}), g({{\Lambda}_{{\bf b},t}^2})$ is
uniform over the square ${[-3, 3n)}^2$ and therefore
$f({{\Lambda}_{{\bf a},t}^1})$ and $g({{\Lambda}_{{\bf b},t}^2})$
considered as random variables are stochastically independent and
themselves have uniform distribution over the appropriate
intervals. Therefore, if the setting $\bf b$ gets changed to the
setting $\bf c$, the random variables $f({{\Lambda}_{{\bf
a},t}^1}), g({{\Lambda}_{{\bf c},t}^2})$ are also independent and
there is no change in the distribution of $f({{\Lambda}_{{\bf
a},t}^1})$ by changing from $\bf b$ to $\bf c$. This should remove
all suspicions of spooky action. We also emphasize that our
construction is highly flexible to introduce uniformity even for
the conditional densities, provided that the time periods
considered are sufficiently long. The method to show this proceeds
along the very same lines as above. It is probably worth noticing
that the sum over all ${\int}ABd{\mu}_i$ resembles a sum over all
possible probability amplitudes in quantum mechanics, except that
everything is real-valued here and dependencies not permitted by
relativity are excluded.

\subsection{Other physical conditions}

To fulfill requirements of physics, it is necessary to be able to
obtain certain values ${-1} \leq {\alpha} \leq 1$ for measurements
on one side only and therefore one needs to be able to have
predetermined values for the following type of integrals
\begin{equation}
{\int}A {\rho} d u d v = \alpha \label{wk15}
\end{equation}
It is easily seen that this can be achieved without changing the
result for the pair correlation by use of  functions $A,B$
generalized in the following way.

\noindent  Define new functions
\begin{equation}
A_r = {A_{r{\bf a}}}(u, z): = {A_{\bf a}}(u) r (z)
\end{equation}
Here $r (z)$ can be any Lebesgue measurable function that assumes
only values $\pm 1$ and  $z$ corresponds to a parameter specific
to the source (e.g. ${{\lambda}_1}, {{\lambda}_2}$ or time $t$).
Similarly define $B_r$ as
\begin{equation}
B_r = {B_{r{\bf b}}}(v, z): = {B_{\bf b}}(v) r (z)
\end{equation}
Then the product $A B = {A_r} {B_r}$, while the integrals of the
single function can be almost arbitrarily adjusted by a proper
choice of $r(z)$. The important special case $\alpha = 0$ is
particularly easy to achieve in a multitude of ways. For example
one can choose a function $r(t)$ (depending only on time $t$) that
varies rapidly and symmetrically between $\pm 1$.

\section{CONCLUSIONS}

We have presented a mathematical framework that can derive the
quantum result for the spin-pair correlation in EPR-type
experiments by use of hidden parameters. A key-element of our
approach is contained in the introduction of time-like statistical
parameters and setting dependent functions of them. This leads in
a natural way to a setting dependent probability measure. The
construction of this probability measure is complicated by the
fact that spooky action must not be introduced indirectly. This is
accomplished by letting the probability measure be a superposition
of SDSPM's with two important properties: (i) the factors of the
product measure depend only on parameters of the station that they
describe and (ii) the joint density of the pairs of setting
dependent parameters in the two stations is uniform. The
mathematical basis for this factorization is the theory of
B-splines.

\section{Acknowledgement}

We would like to thank Anthony J. Leggett for his many
penetrating, critical and very constructive comments on earlier
drafts of this paper. We also would like to thank Gordon A. Baym
and Michael B. Weissman for valuable discussions, Stephen L.
Portnoy for helpful hints regarding B-splines and Richard Blahut
for carefully reading the manuscript. K.H. is grateful to Juergen
Jacumeit, John. R. Barker, David. K. Ferry and Larry R. Cooper for
input in the early stages of this work and acknowledges support of
the Office of Naval Research.

\section{APPENDIX 1}

We show via a simple example that the mixed parameters
${{\Lambda}_{{\bf a},t}^1}$ and ${{\Lambda}_{{\bf b},t}^2}$ cannot
be absorbed in an enlarged parameter space. We introduce unit
vectors $\bf a$ and $\bf b$ as before and define
\begin{equation}
\|{\bf a}\| = |{a_1}| + |{a_2}| + |{a_3}| \label{a1}
\end{equation}
Then
\begin{equation}
1 \leq {\|{\bf a}\|} \leq {3^{1/2}} \label{a2}
\end{equation}
Suppose we have two systems of functions $A = \pm1$ and $B = \pm1$
where $A$ depends only on $\bf a$, $\lambda$ and $\alpha$, and $B$
only on $\bf b$, $\lambda$ and $\beta$ with
$0<\lambda,\alpha,\beta \leq 1$ and such that now $P({\bf a}, {\bf
b})$ is given by
\begin{equation}
P({\bf a}, {\bf b}) = {{\int}_{{(0,1]}^3}} A({\bf a}, \lambda,
\alpha)B({\bf b}, \lambda, \beta)\rho(\lambda, \alpha, \beta)
d{\lambda}d{\alpha}d{\beta} \label{a3}
\end{equation}
and where $\rho$ does not explicitly depend on $\bf a$ and $\bf
b$. Suppose that the parameters $\alpha$ and $\beta$ are allowed
to depend on $\bf a$ and $\bf b$ respectively, just as the
${{\Lambda}_{{\bf a},t}^1}$ and ${{\Lambda}_{{\bf b},t}^2}$ above
are allowed to do. To fix the ideas, let
\begin{equation}
\alpha = {x^{\|a\|}} \text{   ,   } \beta = {y^{\|b\|}} \text{  ,
} 0<x, y \leq 1 \label{a4}
\end{equation}
Then, taking the Jacobian into account, we obtain for the integral
in Eq.(\ref{a3})
\begin{equation}
P({\bf a}, {\bf b}) = {{\int}_{{(0,1]}^3}} A({\bf a}, \lambda,
{x^{\|a\|}})B({\bf b}, \lambda, {y^{\|b\|}})\rho(\lambda,
{x^{\|a\|}}, {y^{\|b\|}})\|a\|\|b\|{x^{\|a\|-1}}{y^{\|b\|-1}}
d{\lambda}dxdy \label{a5}
\end{equation}
Of course, $A$ as well as $B$ can be rewritten as ${A^*}({\bf a},
\delta)$ and ${B^*}({\bf b}, \delta)$ where $\delta$ ranges in an
enlarged parameter set. However, for $\rho$ to be independent of
$\bf a$ and $\bf b$, $\rho$ will have to be a function of very
special form. We conclude that the generalized scenario of station
dependent parameters cannot be handled by Bell's approach.

We would like to add two remarks. First, in the above discussion
we have considered only the case where the parameters $\alpha$ and
$\beta$ do not depend on a common parameter, unlike
${{\Lambda}_{{\bf a},t}^1}$ and ${{\Lambda}_{{\bf b},t}^2}$ which
both depend on $\omega$. The resulting integrals would become line
or surface integrals, making $\rho$ not only dependent on $\bf a$
and $\bf b$ but also on the surface defining these integrals.
Second, we have skirted the issue of properly defining the
integrals corresponding to Eq.(\ref{eq4}), after the mixed
parameters are added in. Since these are stochastic processes
indexed by ${\bf a},t$ these integrals would have to be stochastic
integrals. Since we do not know the precise nature of the mixed
parameters, the question of measurability would be difficult to
address.

\section{APPENDIX 2/LEMMA}

The lemma is an immediate consequence of theorem $4.21$ of
Schumaker \cite{schumaker} for the special values of $m = 3$, $l =
0$, $r = n$, and the knots chosen to be ${y_{\nu}} = \frac {\nu}
{n}$ with $\nu = 0, \pm 1, \pm2,...$. Then by Schumaker's
\cite{schumaker} equation (4.33) we have (dropping the fixed
superscript $3$ of $N_i^3$):
\begin{equation}
{{(y - x)}^2} = {{\sum}_{i=-2}^n}{{\phi}_{i,3}}(y){N_i}(x) \text{
for all } {0  \leq x \leq 1} \text{ and } y \in R \label{wk20}
\end{equation}
Here
\begin{equation}
{{\phi}_{i,3}}(y) = (y - {y_{i+1}})(y - {y_{i+2}})
\end{equation}
and
\begin{equation}
0 \leq {{N_i}(x)} \leq 1 \text{ for all  }x \label{wk21}
\end{equation}
We now restrict $y$ to $0 \leq y \leq 1$. Then for ${-2} \leq i
\leq n$, we have $0 \leq {{{\phi}_{i,3}}(y)} < 2$ unless
$y\in[{y_{i+1}}, {y_{i+2}}]$. Since we must avoid negative $\phi$,
we set $\phi = 0$ in this interval by defining new functions
$\psi$:
\[ {{\psi}_i}(y) = 0 \text{ if } y \in [y_{i+1},
y_{i+2}] \nonumber \]
\[ {{\psi}_i}(y) = {{\phi}_{i,3}}(y) \text { otherwise } \]

Since for $y \in [y_{i+1}, y_{i+2}]$ we have
\begin{equation}
{\mid(y - y_{i+1})(y - y_{i+2})\mid} \leq {\frac {1} {4{n^2}}}
\label{wk22}
\end{equation}
we have
\begin{equation}
0 \leq {{{\sum}_{i=-2}^n}({{\psi}_i}(y){N_i}(x) -
{{\phi}_{i,3}}(y){N_i}(x))} \leq {\frac {1} {4{n^2}}} \label{wk23}
\end{equation}
because for any given $y$ and for all $x$ , only one term in the
sum can be off by at most ${\frac {1} {4{n^2}}}$. This proves the
lemma.

\section{APPENDIX 3}

Assume that the functions $A_{\bf a}$ and $B_{\bf b}$ (considered
as random variables) are independent. Then by Eq.(\ref{eq4})
\begin{equation}
E\{{A_{\bf a}}{B_{\bf b}}\} = E\{{A_{\bf a}}\}E\{{B_{\bf b}}\} =
F({\bf a})G({\bf b}) \label{w2}
\end{equation}
where $F$ and $G$ are functions that depend only on $\bf a$ and
$\bf b$, respectively. But this is in contradiction with
Eq.(\ref{sept1}) since
\begin{equation}
-{\bf a}\cdot{\bf b} = F({\bf a})G({\bf b}) \label{w3}
\end{equation}
and substitution of the three pairs $(1,0,0)/(0,1,0)$,
$(1,0,0)/(1,0,0)$ as well as the pair $(0,1,0)/(0,1,0)$ for ${\bf
a}/{\bf b}$ gives a contradiction.

\end{document}